\documentclass[conference]{IEEEtran}
\IEEEoverridecommandlockouts
\renewcommand{\thetable}{\arabic{table}}
\usepackage{cite}
\usepackage{amsmath,amssymb,amsfonts}
\usepackage{algorithmic}
\usepackage{graphicx}
\usepackage{textcomp}
\usepackage{xcolor}
\usepackage{courier}
\usepackage{subcaption}
\def\BibTeX{{\rm B\kern-.05em{\sc i\kern-.025em b}\kern-.08em
    T\kern-.1667em\lower.7ex\hbox{E}\kern-.125emX}}
\begin{document}

\title{A High Throughput Parallel Hash Table on FPGA using XOR-based Memory}

\author{\IEEEauthorblockN{Ruizhi Zhang, Sasindu Wijeratne, Yang Yang, Sanmukh R. Kuppannagari and Viktor K. Prasanna}
\IEEEauthorblockA{\textit{Department of Electrical and Computer Engineering} \\
University of Southern California, Los Angeles, USA\\
\{ruizhiz, kangaram, yyang172, kuppanna, prasanna\}@usc.edu}
}

\maketitle

\begin{abstract}
Hash table is a fundamental data structure for quick search and retrieval of data. It is a key component in complex graph analytics and AI/ML applications. State-of-the-art parallel hash table implementations either make some simplifying assumptions such as supporting only a subset of hash table operations or employ optimizations that lead to performance that is highly data dependent and in the worst case can be similar to a sequential implementation. In contrast, in this work we develop a dynamic hash table that supports all the hash table queries - search, insert, delete, update, while allowing us to support $p$ parallel queries ($p>1$) per clock cycle via $p$ processing engines (PEs) in the worst case i.e. the performance is data agnostic. We achieve this by implementing novel XOR based multi-ported block memories on FPGAs. Additionally, we develop a technique to optimize the memory requirement of the hash table if the ratio of search to insert/update/delete queries is known beforehand. We implement our design on state-of-the-art FPGA devices. Our design is scalable to 16 PEs and supports throughput up to 5926 MOPS. It matches the throughput of the state-of-the-art hash table design -- FASTHash, which only supports search and insert operations. Comparing with the best FPGA design that supports the same set of operations, our hash table achieves up to 12.3$\times$ speedup.
\end{abstract}


\section{Introduction}
Hash table is a key data structure for applications requiring quick search and retrieval. Applications such as graph analytics, AI/ML~\cite{boulis2005text,vijayanarasimhan2018large,holt2002mining,wang2018supervised,zeng2019accurate,slide} widely use hash tables. For example, the graph sampling operation of Graph Convolution Neural Network (GCN) uses hash tables to determine whether a vertex (or edge) currently sampled exists in the sampled set of not~\cite{zeng2019accurate}. Similarly, hash tables are used to quickly select the nearest neighbors in Approximate Nearest Neighbor (ANN) algorithm~\cite{wang2018supervised} and to maintain bag-of-words in text mining applications~\cite{boulis2005text}.

Field Programmable Gate Arrays (FPGA) have found widespread success in accelerating data and compute intensive applications due to the availability of fine grained parallelism~\cite{kuppannagari2014energy}. They have dense reconfigurable logic elements (up to 5.5 million) and large on-chip memory (up to 500 Mb) allowing users to create highly optimized designs to achieve high performance~\cite{xilinxalveo,FPGA-Altera-Stratix10}. FPGAs are widely deployed in high performance cloud and data-centre platforms~\cite{xilinxalveo} as well as in low-powered edge applications~\cite{hao2019fpga}. 

Several works have focused on increasing the throughput of hash table implementations by ``parallelizing" i.e. by exploiting certain features of the hash table, such as the availability of multiple partitions~\cite{fpga_cuckoo_parallel}. However, the parallelism that can be obtained using such approaches is highly data dependent and the worst case performance - for example, when all queries belong to the same partition -  is similar to a serial implementation. 

In~\cite{isc2020}, we developed a hash table that supports $p$ queries ($p > 1$) in each clock cycle, where $p$ is the number of parallel Processing Engines (PEs), even in the worst case. In other words, the performance of the hash table is agnostic to the data. Such a guarantee was obtained by making the assumption that the application using the hash table requires only search and insert operations. 

In this work, we address the limitations of~\cite{isc2020} by completely redesigning the hash table to support all hash table operations - search, insert, update and delete. To achieve the same, we implement XOR based multi-ported block memories to be used as storage for the hash table, albeit, at the cost of increased memory requirement. Additionally, we develop a technique to reduce the memory requirements of the hash table if the ratio of read-only (search) and read-write (insert, update, delete) operations are known beforehand. Our hash table supports the same relaxed consistency semantics as the design in~\cite{isc2020}. 

The key contributions of the paper are as follows:
\begin{itemize}
    \item We develop a dynamic parallel hash table table by implementing XOR-based multi-ported block memories.
    \item Our XOR-based block memories fully utilize the abundant on-chip SRAM to support $p$ queries - search, insert, update, delete ($p > 1$) in each clock cycle even in the worst case.   
    \item Our hash table architecture allows customization if workloads' non-search queries per cycle has an upper-bound, thereby further reducing the memory resource consumption.
    \item We implement the hash table on state-of-the-art FPGAs and show that our hash table supports 16 parallel queries per cycle reaching a throughput of 5926 million operations per second at 370 MHz. 
\end{itemize}

\section{Related Work}
Several works have focused on FPGA-based high performance hash table implementations. In~\cite{fpga_route_lookup}, the authors proposed a hash table based IP lookup technique that achieves a lookup throughput of 250 Mops/s. Authors in~\cite{fpga_bloom_filter} developed a hash table implementation using bloom filter which reduces unnecessary hash table accesses. In~\cite{fpga_da_tong}, the authors developed a off-chip DRAM based hash table. To avoid data hazards the authors developed a forwarding unit for dynamic hash table updates. Works such as~\cite{fpga_cuckoo_memory_efficient} and~\cite{fpga_cuckoo_parallel} have developed efficient hash tables using cuckoo hashing scheme. In~\cite{fpga_cuckoo_memory_efficient} the authors incorporate a decoupled key-value storage to enable parallel computation of hash values. In~\cite{fpga_cuckoo_parallel}, the authors developed a Cuckoo hash table with multiple parallel pipelines with each pipeline having a different entry point. 


The limitation of the current works is as following: (i) They either focus on improving performance for a single processing pipeline, which is not sufficient to fully exploit the high bandwidth on-chip SRAM in state-of-the-art FPGAs. Examples include~\cite{fpga_route_lookup,fpga_bloom_filter,fpga_kvs_10gbps,fpga_da_tong,fpga_cuckoo_memory_efficient} or (ii) their parallelism is hash table or data dependent. For example,  the parallelism of~\cite{fpga_cuckoo_parallel} is limited by both the number of hashing functions in a given Cuckoo hash table and the data conflicts when keys-value being accessed by different pipelines fall into the same partition.

In addition, several works have focus on developing parallel hash tables for CPU and GPU platforms. On CPU, the focus has been on developing concurrent and lock-free hash table through shared memory and message passing~\cite{multicore_Metreveli,multicore_shalev,multicore_michael}. A recent work~\cite{simdhash} developed a hash table implementation using SIMD extensions. However, the hash table is only limited to lookups. On GPU, the focus has been on dividing a hash table into coarse or fine grained partitions and extracting parallelism by concurrent processing~\cite{stadium_hashing,owen_hash_table,robinhood_hashing}.

\textbf{Novelty of Our Work:} State-of-the-art works either improve the performance of a single pipeline by increased pipelining to increase clock frequency or they perform parallel processing by atomic accesses of hash table partitions. The former approach is severely limited in achievable throughput while the latter is highly data dependent. This is because in the worst case the parallel queries might need to access the same portion of the hash table leading to serialized processing within partitions. To address these limitations, in our previous work~\cite{isc2020}, we developed a hash table that supports $p$ queries ($p > 1$) in each clock cycle via $p$ parallel Processing Engines, even in the worst case. In other words, the performance of the hash table is agnostic to the data. Such a guarantee was obtained by making the assumption that the application using the hash table requires only search and insert operations. In this work, we improve upon the design by supporting \underline{all} common hash table operations.

\section{Hash Table Definition and Supported Operations}

Our design is a standard closed addressing hash table. Each hash table bucket has multiple slots for collision resolution, each of which can be used to store one key-value pair. A key-value pair with key $x$ is placed in index \textit{h(x)}, where $h(\cdot)$ is the hash function used by the hash table. Our hash table supports the following operations:

\begin{itemize}
\item Search ($key$): Search operation retrieves the $value$ associated with $key$ if $key$ exists in the hash table. Otherwise returns $None$.
\item Insert/Update ($key$, $value$): Insert/Update operation first searches $key$ in the hash table. If $key$ exists already, it updates the old value with $value$. Otherwise, a new $key$-$value$ pair is inserted into an open slot of the hash table.
\item Delete ($key$): Delete operations removes the $key$-$value$ pair from the hash table, and marks the corresponding slot as open.

\end{itemize}

In this paper, we refer hash table mutation queries - Insert, Update, and Delete - as Non-Search Queries (NSQ). A parallel hash table is able to process queries concurrently. The throughput of parallel hash table often suffers from contentions between concurrent queries. The extent these problems manifest depends on individual design. Previous works often divide hash table into partitions, and extract parallelism by accessing each partition concurrently. However, when queries collide and access the same partition, either fine-grained~\cite{owen_hash_table} or coarse-grained~\cite{multicore_Metreveli}, they are intrinsically serialized. Unlike previous designs, our parallel hash table guarantees $p$ parallel accesses per cycle in the worst case. We further define NSQ ratio as follows:

\newtheorem{nsq_def}{Definition}
\begin{nsq_def}
\normalfont
The non-search queries (NSQ) ratio of a parallel hash table that can process $p$ queries per cycle is defined as the ratio of the maximum number of NSQ it can receive in a single cycle -- $k$ and the maximum number queries it can process per cycle -- $p$. Thus, NSQ ratio = $k/p$.
\end{nsq_def}

\section{Hash Table on FPGA}

\subsection{Challenges in Parallel Hash Table}

The goal of our parallel hash table is to guarantee $p$ search, insert, update, or delete queries per cycle via $p$ PEs. A parallel memory with $p$ read ports and $p$ write ports is needed to sustain the throughput requirements. Since each FPGA SRAM block only has 1 read and 1 write port, it is non-trivial to support all the operations without stalling the pipelines. ~\cite{isc2020} introduces a parallel architecture that supports search and insert operations with throughput guarantee. To support update and delete in their design, it requires re-routing update and delete queries to the PE which receives the  original insert for the corresponding input key. Besides the extra logic complexity increase (all-to-all PE connections), throughput of the hash table is no longer guaranteed, and can drop based on access patterns. To overcome these challenges, our hash table architecture utilizes XOR-based block memories to support search, insert, update and delete queries with $p$ queries per cycle guaranteed throughput.

\subsection{XOR-based Block Memory}

\begin{figure}[ht]
\vspace{-0.3cm}
\includegraphics[width=1\linewidth]{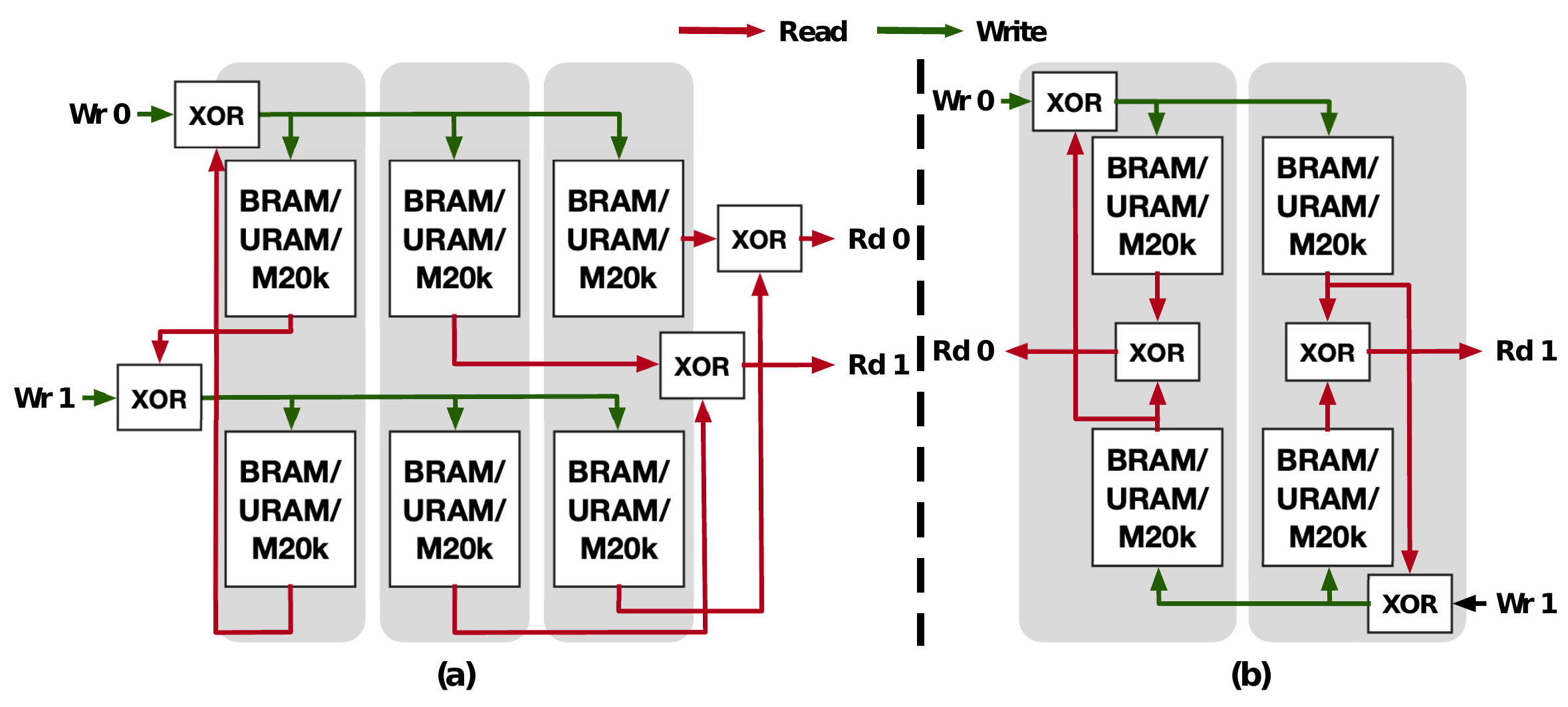}
\centering
\caption{An example design of a 2R2W XOR-based memory. (a): Original design in~\cite{xor-sram}. (b): Design used in our architecture.} 
\vspace{-0.2cm}
\label{xor-sram}
\end{figure}

LaForest et al~\cite{xor-sram} introduced XOR-based multi-ported memory using FPGA on-chip SRAM blocks (BRAM, URAM, M20K). The key property of \texttt{XOR} is that using the \texttt{XOR} value of A and B, one can recover A by \texttt{XOR} this value with B. As a result, each SRAM block inside an XOR-based multi-ported memory stores the \texttt{XOR}'ed data, instead of the actual data. It is shown that XOR-based memory can efficiently scale to large number of read and write ports on FPGA, yet doesn't incur significant logic complexity as the other approaches~\cite{xor-sram}. Figure~\ref{xor-sram}(a) shows an example with 2R2W XOR-based memory. To perform a read, data in each SRAM block (third shadowed column) is read out and \texttt{XOR}'ed. For write, the new data needs to be \texttt{XOR}'ed with the data in the SRAM block first (a read followed by \texttt{XOR}), then the \texttt{XOR}'ed data can be written to each SRAM block. More details on this technique can be found in~\cite{xor-sram,xor-mem-2}.

A drawback of the XOR-based approach is that it consumes extra SRAM resources -- a $m$R$n$W XOR-based memory requires $n\times$($n$-1+$m$) SRAM blocks~\cite{xor-sram}. To reduce memory resource consumption, read and write operations of the XOR-based memory in our design share the same read ports. As a result, the memory requirement is reduced to $m$$\times$$n$. Figure~\ref{xor-sram}(b) shows our memory design for 2R2W.

\subsection{Hash Table Architecture}

Our proposed design is composed by an array of processing engines (PEs). A hash table is replicated $p$ times, and each PE holds one replica. Inside each PE, key-value pairs are stored as XOR-encoded data across $k$ \textbf{Partial XOR Store}, where $k$ is the max number of non-search queries the architecture can handle per cycle. To handle hash table collision, each Partial XOR Store has multiple slots (2 - 4 being the common case), and each slot stores one key-value pair.



\begin{figure}[ht]
\vspace{-0.3cm}
\includegraphics[width=1\linewidth]{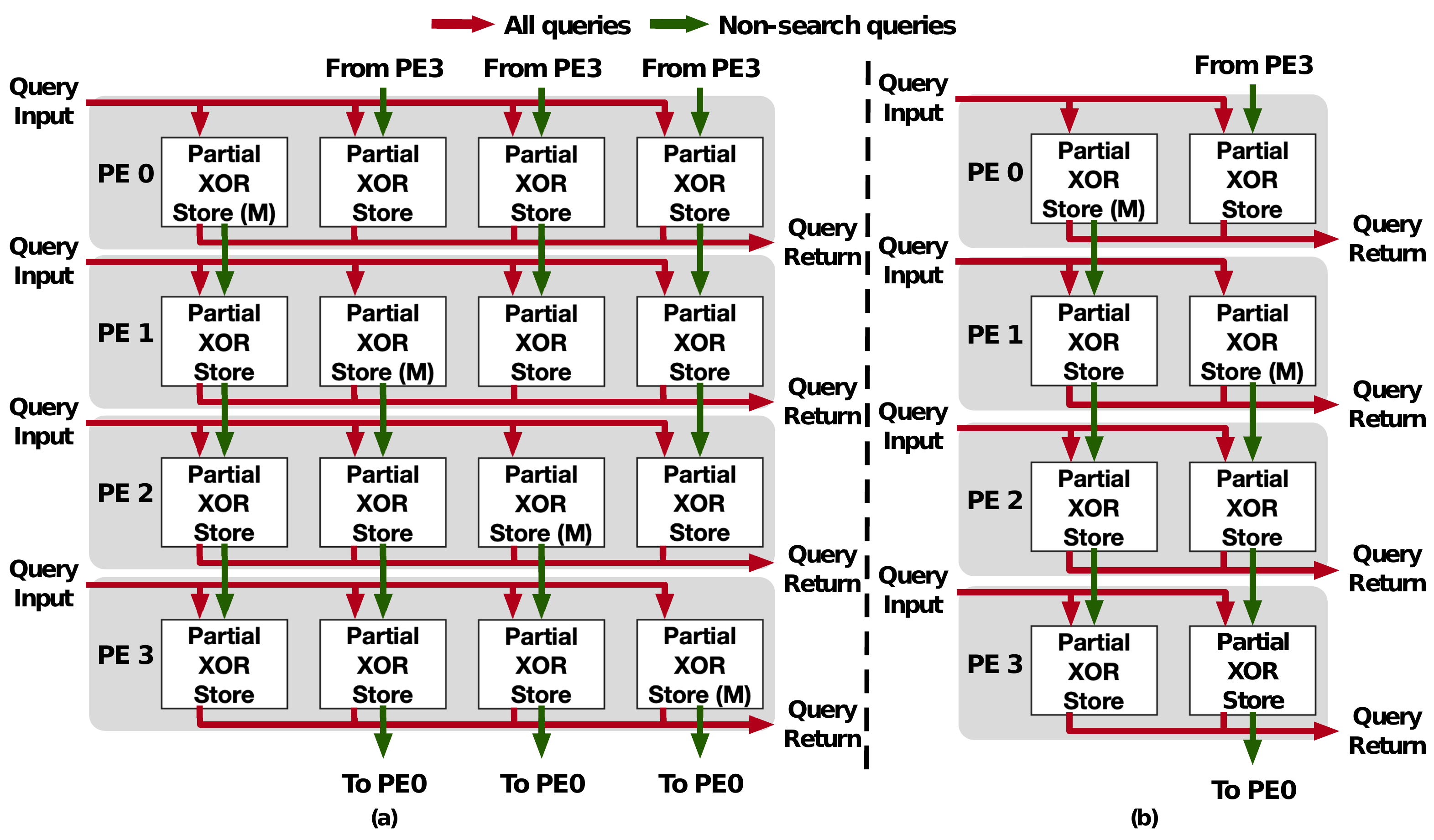}
\centering
\caption{Query dataflow of a 4 PEs design. (a): A configuration that supports up to 100\% non-search queries per cycle. (b): A configuration that supports up to 50\% non-search queries per cycle.} 
\label{sys}
\vspace{-0.4cm}
\end{figure}

\subsubsection{Query Dataflow}

NSQ initiated by one replica need to be updated by all the other replicas in order to ensure data consistency. To enable conflict-free synchronization between replicas, we design a special component called \textbf{Partial XOR Store (M)}. This component only exists in PEs that can process hash table mutation operations (insert, update, and delete). It is the initiation point where mutation operations start and connects to Partial XOR Store in the other PEs as a pipeline. With this component, our architecture guarantees mutations performed by one PE never have conflict with other PEs.

The query flow of our hash table, i.e. mapping of search, insert, update, and delete operations to our parallel architecture, is as follows:

$Search$: Once a PE receives search query, it first calculates the bucket index using a predefined hash function. Then, PE initiates read to all the Partial XOR Store in parallel using the calculated index. If all partial stores indicate valid data, an XOR operation is performed to recover the original key-value pair by XOR'ing the data from each Partial XOR Store together. PE returns the key-value pair back to the application when there is a match.

$Insert/Update$: Insert/Update queries first go through the same dataflow as Search queries to decide if the key-value pair exists or not. Following that, PE generates the XOR-encoded data, which will be written to the Partial XOR Store (M) first. For new key-value pair, the encoded data is the same as the original data; for updates, PE generates encoded data by XORing the incoming key-value pair with existing data from each Partial XOR Store (note that this excludes the encoded-data in Partial XOR Store (M)). New key-value pair also needs to have an open slot in order to proceed (collision handling). After that, $p$ subsequent writes with the same encoded data is sent to remote PEs via inter-PE communication.

$Delete$: Delete queries also go through the same data flow as Search queries first. If there is a match to the input key, PE sets the invalid bit in the corresponding location in Partial XOR Store (M). This invalidation also propagates to other PEs like Insert/Update queries.

\textbf{Workloads Specific Customization}: The memory requirements of our architecture increases as $p$ and $k$ increase. Our design allows customization when non-search query (NSQ) per cycle of a workload has an upper-bound. Search-only PEs without hash table mutation capability will be used to save memory resource consumption. The key difference between search-only PEs and full PEs is that there is no Partial XOR Store (M)) component in search-only PEs (details in next subsection). As an example, Figure~\ref{sys}(a) shows 4 PEs query dataflow with $p$=4 and $k$=4. Figure~\ref{sys}(b) illustrates a design with $p$=4 and $k$=2.

\subsubsection{Processing Engine Design}

\begin{figure}[ht]
\vspace{-0.3cm}
\includegraphics[width=0.8\linewidth]{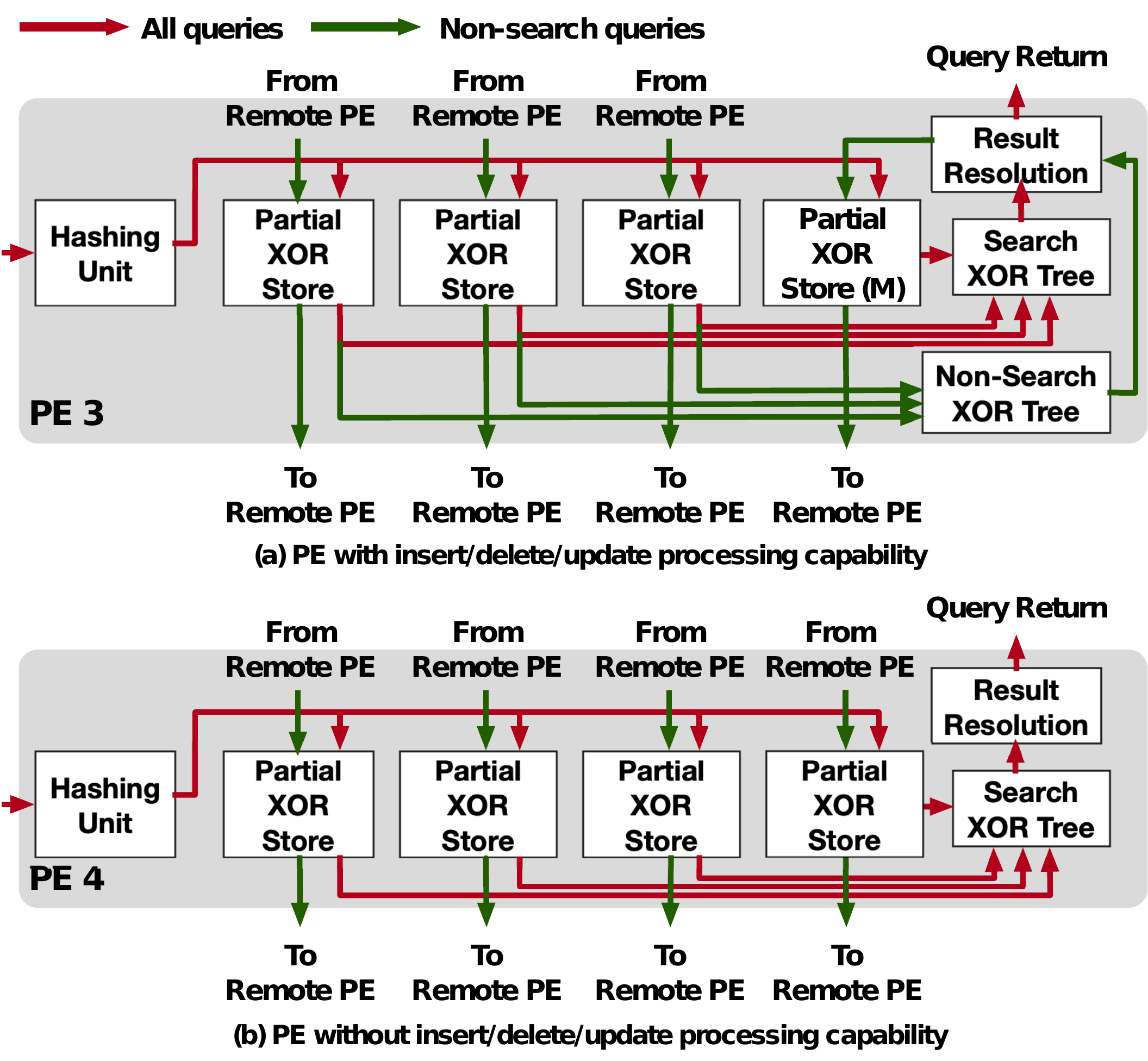}
\centering
\caption{Architecture of PEs in an 8-PE design that supports up to 50\% non-search query per cycle. (a) PE 3 can process all types of queries. (b) PE 4 can only process search queries.} 
\label{pe}
\vspace{-0.3cm}
\end{figure}

Figure~\ref{pe}(a) shows the pipeline of our PE with full processing capability. The key components are hashing unit, Partial XOR Store, XOR reduction tree, and result resolution unit. Each component processes query in one or more cycles. Hashing unit receives queries and calculate bucket index. We use the Class $H_{3}$ \cite{h3} hashing, which has been demonstrated to be effective on distributing keys randomly among hash table entries. The hash function is defined as follows \cite{h3_2}:

\newtheorem{h3_def}{Definition}
\begin{h3_def}
\normalfont
For key of bitwidth $i$ and hash index of bitwidth $j$, let $Q$ denote a $i$ $\times$ $j$ Boolean matrix. For a given $q$ $\in$ $Q$, let $q$($m$) be the bit string of the $m$th row of $Q$, and let $x$($m$) denote the $m$th bit of input key. The hash function is: $h$($x$) = ($x$($1$) $\cdot$ $q$($1$)) $\oplus$ ($x$($2$) $\cdot$ $q$($2$)) $\oplus$ ... $\oplus$ ($x$($i$) $\cdot$ $q$($i$)).
\end{h3_def}

As discussed in query dataflow, once the bucket index is calculated, PE initiates read operation to each Partial XOR Store in parallel. The XOR-encoded data are read out and fed into two XOR reduction trees. The first XOR reduction tree is to recover the original key-value pair. The second XOR reduction tree is only active for non-search queries and is used to produce new encoded data. 

Result resolution unit collects the decoded key-value pair from XOR reduction tree, and route query to the next hop based on operation type and query dataflow. It also performs collision resolution by probing each slot in parallel and finding the first open slot.

Our flexible architecture allows customization for workloads with known NSQ ratio to reduce SRAM resource consumption. Figure~\ref{pe}(b) depicts the architecture of PE that only support search queries. Both Partial XOR Store \textbf{(M)} and Non-Search XOR tree are removed as they are designed to enable non-search queries processing capability.

\subsubsection{Inter-PE Communication}

Inter-PE data communication plays an important role for the consistency of hash table replicas. It is a pipeline across different PEs -- NSQ from one PE takes $p$ cycles to fully propagates to all other PEs. As illustrated in Figure~\ref{sys}, all hash table mutations starts from Partial XOR Store (M). Since each PE has at most one Partial XOR Store (M) and they do not share common datapath. This communication mechanism ensures mutations initiated by one PE never collide with mutations initiated from other PEs.

\subsection{Analysis of Memory Requirements}

On-chip SRAM block is the main resource consumption for our architecture, due to the hash table replications and XOR-encoded data store. Multiple factors can influence the SRAM consumption: 1) Target throughput or the number of PEs. 2) Max NSQ per cycle of an implementation. 3) Key-value pair size 4) Collision handling capability or slots per hash table entry. Hence, an optimal design should consider all these factors, as well as workload characteristics and device resource availability. 

\begin{figure}[ht]
\vspace{-0.5cm}
\includegraphics[width=1\linewidth]{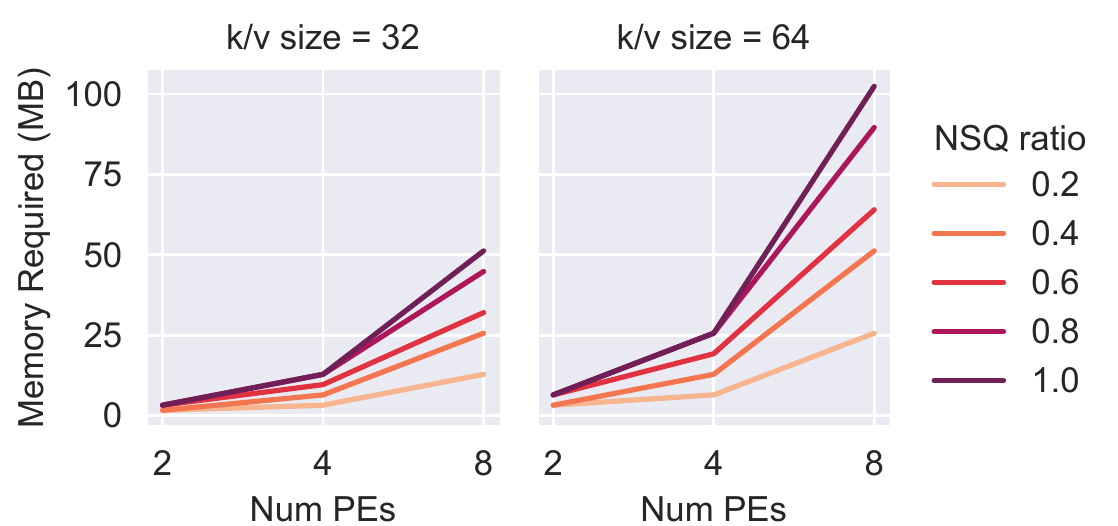}
\centering
\caption{Memory requirements for different PE configurations. Hash table capacity is fixed at 50k entries and 2 slots per entry.} 
\label{memory}
\vspace{-0.3cm}
\end{figure}

Figure~\ref{memory} illustrates the SRAM resource requirements for a hash table with 50K entries and 2 slots per entry. The key and value sizes are both 4 bytes. Given different hash table configurations (number of PEs, ratio defined as $k$/$p$), the SRAM requirements diff significantly, ranging from less than 2 MB to more than 100 MB. As a reference point, SOTA FPGAs such as Xilinx U250 has around 50 MB on-chip SRAM~\cite{xilinxalveo}.

\subsection{Hash Table Consistency Model}

As described in Section III, a new key is not seen by all the PEs for up to $p+t_0$ cycles. $t_0$ includes hashing calculation, partial XOR read, and result resolution. It is a constant because our design performs parallel read to all the blocks. In addition, $p$ cycles is needed to perform hash table mutation to all the PEs through Inter-PE communication.

In~\cite{isc2020}, authors define the notion of relaxing the consistency of hash table. The key idea is that by not using complicated forwarding units to address RAW and WAW hazards, inconsistency in hash table may occur when: (i) a NSQ for a key $u$ is received by one PE; and (ii) another query for the same key $u$ is received by another PE while the first one is still being processed. However, the maximum number of such erroneous queries $n_{err}$ is bounded, as each cycle the architecture can process $p$ queries, and the latency for each query type is within $p+t_0$ cycles. Therefore, following the same proof as ~\cite{isc2020} (omitted due to space limitation), our hash table architecture has the following consistency guarantee:

\textbf{Theorem 1.} Number of queries incorrectly served due to relaxed consistency is given by $P(n_{err} \geq \theta) \leq \frac{p^2+pt_0}{\theta}$.

\section{Experimental Evaluation}

\subsection{Experimental Setup}

We implemented the hash table design on the Xilinx Alveo U250~\cite{xilinxalveo} as well as on the Intel Stratix 10 GX1800 FPGA~\cite{intelstratix} using Verilog HDL. The Xilinx device has 1,728,000 LUTs, 3,456,000 Flip-Flops, and 360 Mb of URAM memory, while the Intel device has 933,120 ALMs, 3,732,480 ALM registers, and 229 Mb of M20K memory. Post-place-and-route simulations were performed using Xilinx Vivado Design Suite 2018.3 and Intel Quartus Prime 19.4, respectively. 

Our modular design and flexible configurability give users a wide range of design options based on application requirements and available FPGA resources. We evaluated its performance and resource utilization by increasing the number of PEs for different non-search queries per cycle ratios and varying the total number of hash entries as well as key-value lengths (k/v lengths). The key/value sizes in our experiments were 32, 64, and 128 bits. These numbers cover the most configurations in graph analytics and AI applications. We generated uniformly distributed access patterns, which include both the operation types and the hash keys, as our stimulus. We used million operations per second (MOPS) as the metric to calculate the throughput of the above configurations. The utilization of FPGA resources is reported in terms of percent usage of LUTs (ALMs), and on-chip memory (SRAM).

\subsection{Evaluation on Xilinx Alveo U250 FPGA}

\begin{figure}[ht!] 
\centering
\includegraphics[width=0.8\linewidth]{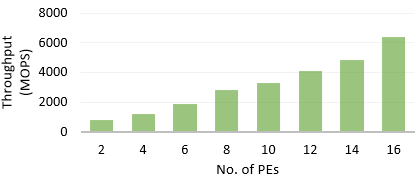}
\caption{Evaluation of hash table on Xilinx device (64-bit key \& value in size)}
\label{virtex_throuput}
\vspace{-0.3cm}
\end{figure}

Figure \ref{virtex_throuput} shows the maximum achievable throughput for various PE counts in the Xilinx Alveo 250U FPGA. The results verify the scalability of our design with the number of PEs. The design's clock rate is in the range of 286 MHz – 400 MHz. We observed high clock frequencies even in a large number of PEs when the design parameters (k/v size, NSQ ratio) well fit on chip even though the pressure on the place and route increases with the number of PEs. Insert latency is higher than search latency due to the extra clock cycles that are spent writing a new key-value pair to all PEs. The search operation and the insert operation can be completed within 14 ns and 54 ns with 16 PEs.

\begin{figure}[ht]
\vspace{-0.3cm}
\begin{subfigure}{.5\textwidth}
  \centering
  \includegraphics[width=0.8\linewidth]{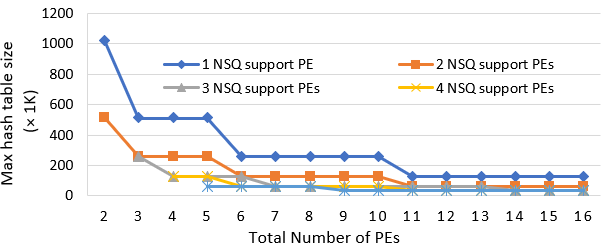}  
  \caption{2 Slots per Entry}
  \label{figsub-figsub-fig_max_entries_xilinx_1}
\end{subfigure}
\begin{subfigure}{.5\textwidth}
  \centering
  \includegraphics[width=0.8\linewidth]{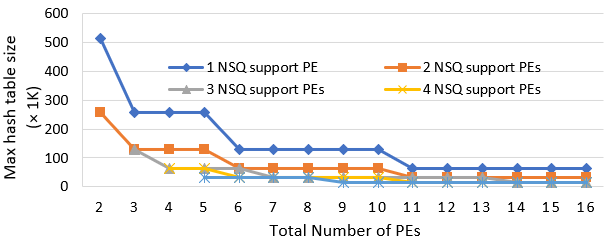}  
  \caption{4 Slots per Entry}
  \label{figsub-fig_max_entries_xilinx_2}
\end{subfigure}
\caption{Max hash table sizes supported on Xilinx U250 FPGA (k/v size = 64-bit)}
\vspace{-0.4cm}
\label{fig_max_entries_xilinx}
\end{figure}

Figure \ref{fig_max_entries_xilinx} shows the fluctuation of the maximum hash table size - the number of entries that can be implemented on the Xilinx U250 FPGA, with 64-bit key and value sizes as the number of PEs increases. We observe the impact of the number of NSQ-supported PEs on the number of entries. The total number of plausible entries drops as the NSQ-supported PEs increased because NSQ-supported PEs utilize twice the URAMs compared to a typical PE.

The resource utilization of a hash table implementation is reported in Table \ref{table-util-xilinx}. The number of entries in a hash table varies from 16K – 128K for different NSQ ratios. Further, they include 4 slots per entry with the k/v length of 64-bit. We make heavy use of URAM for the on-chip hash table store.

\begin{table}[ht!] 
\centering
\caption{Resource Utilization of Dynamic Hash Table on Xilinx U250 FPGA (Number of Slots = 4)}
\label{table-util-xilinx}
\begin{tabular}{|c|c|c|c|c|c|}\hline
\textbf{\# of Entries} & \textbf{\# of PEs} & \textbf{NSQ PE ratio} & \textbf{LUT} (\%) & \textbf{URAM (\%)} \\\hline\hline
128K & 4 & 2/4 & $<$ 1\% & 80\% \\\hline
64K & 8 & 2/8 & $<$ 1\% & 80\% \\\hline
32K & 16 & 2/16 & $<$ 1\% & 80\% \\\hline
16K & 7 & 8/8 & $<$ 1\% & 80\% \\\hline
\end{tabular}
\vspace{-0.1cm}
\end{table}

\begin{figure}[ht!] 
\vspace{-0.3cm}
\centering
\includegraphics[width=0.8\linewidth]{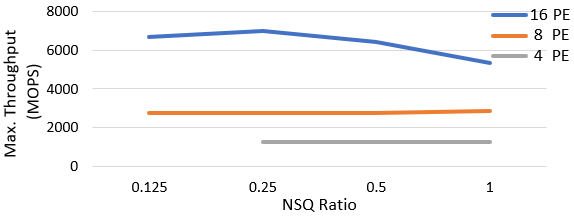}
\caption{Maximum throughput fluctuation w.r.t NSQ ratio}
\label{max_throuput_uo}
\vspace{-0.2cm}
\end{figure}

Figure \ref{max_throuput_uo} shows the effect of the supported non-search queries per cycle on maximum throughput. For the same number of PEs, some configurations show high throughput as their arrangement fits well on the FPGA, which leads to high clock frequency.

\subsection{Evaluation on Intel Stratix 10 FPGA}

\begin{table}[ht!] 
\vspace{-0.2cm}
\centering
\caption{Resource Utilization of Dynamic Hash Table on Stratix 10 FPGA (Number of Slots = 4 and 64-bit key \& value in size)}
\label{table-util-intel}
\begin{tabular}{|c|c|c|c|c|c|}\hline
\textbf{\# of Entries} & \textbf{\# of PEs} & \textbf{NSQ PE ratio} & \textbf{LUT} (\%) & \textbf{URAM (\%)} \\\hline\hline
128K & 2 & 2/2 & $<$ 1\% & 57\% \\\hline
64K & 4 & 2/4 & 1\% & 56\% \\\hline
32K & 6 & 2/6 & 1\% & 42\% \\\hline
16K & 8 & 4/8 & 1\% & 56\% \\\hline
\end{tabular}
\vspace{-0.2cm}
\end{table}

\begin{table*}[ht] 
\centering
\caption{Existing hash table implementations compared to proposed design}
\label{tablecomparefpga}
\begin{tabular}{|c|c|c|c|c|c|}\hline
\textbf{Metric} & \textbf{This work} & \textbf{Yang~\cite{isc2020}} & \textbf{Pontarelli~\cite{fpga_cuckoo_parallel}} & \textbf{Ashkiani~\cite{owen_hash_table}} & \textbf{Awad~\cite{ipdpstable}} \\\hline\hline
Technology & FPGA & FPGA & FPGA & GPU & GPU \\\hline
Max. Throughput & 5926 & 5360 & 480 & 937 & 1015 \\
(Search MOPS) & & & (Normalized Fmax) & & \\\hline
Supported Queries$^1$ & S, I, U, D & S, I & S, I, U, D & S, I, U, D & S, I, U, D \\\hline
\end{tabular}
\vspace{-0.5cm}
\end{table*}

\begin{figure}[ht!] 
\vspace{-0.2cm}
\centering
\includegraphics[width=0.7\linewidth]{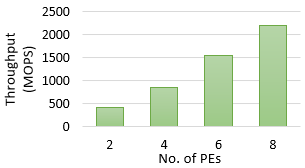}
\caption{Evaluation of hash table on Intel device (64-bit key \& value in size)}
\label{stratix_throuput}
\vspace{-0.2cm}
\end{figure}

We also implemented our hash table on Intel Stratix 10 GX2800 FPGA. We used 32-bit and 64-bit as k/v size and used 2/4 slots per entry. Figure \ref{stratix_throuput} shows the maximum achievable throughput for various PE counts in the Intel Stratix 10 FPGA. The result indicates that our architecture is designed as a general solution to work with various FPGA devices. This design can process up to 2208 MOPS with 8 PEs. The number of PEs affects the max clock rate; the achieved throughput does not scale linearly.

\begin{figure}[ht]
\begin{subfigure}{.5\textwidth}
  \centering
  \includegraphics[width=0.8\linewidth]{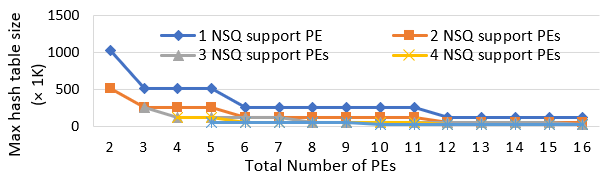}  
  \caption{2 Slots per Entry}
  \label{figsub-figsub-fig_max_entries_intel_1}
\end{subfigure}
\begin{subfigure}{.5\textwidth}
  \centering
  \includegraphics[width=0.8\linewidth]{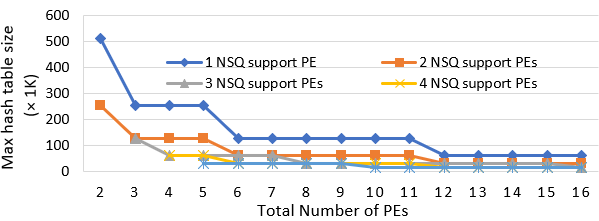}  
  \caption{4 Slots per Entry}
  \label{figsub-fig_max_entries_intel_2}
\end{subfigure}
\caption{Max hash table sizes supported on Intel Stratix 10 FPGA (k/v size = 32-bit)}
\vspace{-0.3cm}
\label{fig_max_entries_intel}
\end{figure}

From Table \ref{table-util-intel}, we can see that the usage of ALMs and registers is low, while the utilization of M20K depends on the hash table capacity and number of PEs. Figure \ref{fig_max_entries_intel} shows the fluctuations of the maximum hash table size - the number of entries that can be implemented on Intel Stratix 10 FPGA, with 32-bit k/v sizes as the number of PEs increases.

\subsection{Comparison with state-of-the-art (SOTA)}

We compare the performance of our 16-PE hash table implementation on Xilinx U250 FPGA with state-of-the-art GPU and FPGA designs. The performance metric is in terms of throughput - MOPS.

The implementation of Yang et al.~\cite{isc2020} uses the same FPGA, namely, the Xilinx Alveo U250. Their implementation operates at 335 MHz with 16 PEs. Pontarelli et al.~\cite{fpga_cuckoo_parallel} propose a parallel Cuckoo hashing on FPGA. The hash table is stored completely on chip. The target FPGA device is the Xilinx Virtex5 XC5VLX155T. Their implementation operates at 156.25 MHz. Awad et al.~\cite{ipdpstable} is implemented on the NVIDIA Titan V (Volta) GPU. This GPU has 5120 CUDA cores that operate at 1455 MHz. In Ashkiani et al.~\cite{owen_hash_table}, the design is implemented on the NVIDIA Tesla K40c GPU. It has 2880 CUDA cores and operates at 745 MHz.

\begin{figure}[ht!] 
\vspace{-0.2cm}
\centering
\includegraphics[width=0.8\linewidth]{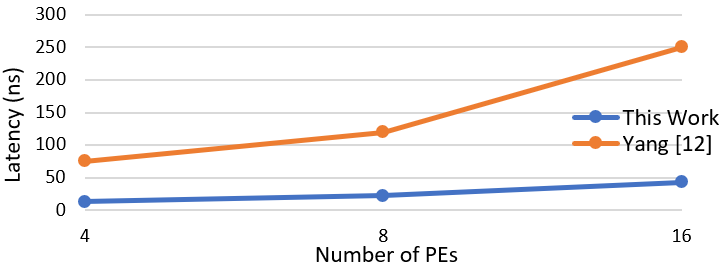}
\caption{Latency comparison with SOTA [Lower the better]} 
\label{latency_comp_sota}
\vspace{-0.4cm}
\end{figure}

Table \ref{tablecomparefpga} summarizes the SOTA designs, and their performance with respect to our work\footnote{For supported queries, S: Search, I: Insert, U: Update, D: Delete.}. Similar to Ashkiani et al.~\cite{owen_hash_table}, we used a 32-bit key/value size and random traffic patterns in the comparison. Compared with the SOTA GPU work, we observe a speedup of 5.8x even while running with 3.8x less clock frequency. Compared with the SOTA FPGA work, our design achieves up to 1.1x and 12.3x speedup. Note that although the max throughput is only 10\% higher than Yang et al~\cite{isc2020}, our architecture supports all four common query types with worst-case throughput guarantee. Further, the latency of our design is significantly lower than Yang et al.~\cite{isc2020}, as shown in Figure \ref{latency_comp_sota}. 

\section{Conclusion}

In this paper, we proposed a high throughput parallel hash table using XOR-based memory. It supports all common hash table operations -- search, insert, update, and delete, while still guarantees a throughput of $p$ queries per cycle in the worst case. We further optimized our hash table architecture to reduce SRAM resource consumption for workloads with different NSQ ratio. Our hash table uses novel query flow and inter-PE communication to ensure data consistency. We implement our design on SOTA FPGA devices, experimental results show that our architecture is flexible and scalable with respect to number of PEs, key/value sizes, number of slots per entry, as well as hash table capacity. Our design achieves up to 5926 MOPS throughput with 16 PEs, matching the throughput of the SOTA hash table design -- FASTHash, which only supports search and insert operations (1.1$\times$ speedup). Comparing with the best FPGA design that supports the same set of operations, our hash table achieves up to 12.3$\times$ speedup.

\section{Acknowledgement}

This work has been supported by the U.S. National Science Foundation (NSF) under grants OAC-1911229 and SPX-1919289.

\bibliographystyle{IEEEtran}
\bibliography{hpec20}
\end{document}